\documentstyle[twocolumn,psfig,aps]{revtex}
\begin{document}
\draft
\twocolumn[\hsize\textwidth\columnwidth\hsize\csname @twocolumnfalse\endcsname
%
%
%

\title{ Indications of Spin-Charge Separation at Short Distance and Stripe Formation \\
in the Extended t-J Model on Ladders and Planes}

\author{G. B. Martins$^1$, J. C. Xavier$^1$, C. Gazza$^2$,  
M. Vojta$^3$, and E. Dagotto$^1$}

\address{$^1$ National High Magnetic Field Lab, Department of Physics, \\
and MARTECH, Florida State University, Tallahassee, FL 32306, USA}

\address{$^2$ Instituto de F\'{\i}sica Rosario (CONICET) and Universidad Nacional 
de Rosario, 2000 Rosario, Argentina}

\address{$^3$ Department of Physics, Yale University,
P.O.Box 208120, New Haven, CT 06520-8120, USA}

\date{\today}
\maketitle

\begin{abstract}
The recently discussed tendency of holes to generate nontrivial
spin environments in the extended two-dimensional
t-J model (G. Martins, R. Eder, and E. Dagotto,
Phys. Rev. B{\bf 60}, R3716 (1999)) is here investigated using
computational techniques applied to
ladders with several number of legs. This
tendency is studied  also 
with the help of analytic spin-polaron approaches directly in two
dimensions. Our main result is that the 
presence of robust antiferromagnetic correlations between spins 
located at both sides of a hole either along the x or y axis, 
observed before numerically on square clusters, is also found
using ladders, as well as applying 
techniques based on a string-basis expansion. 
This so-called ``across-the-hole'' 
 nontrivial structure exists even in the two-leg spin-gapped 
ladder system, and leads to an effective reduction in dimensionality 
and spin-charge separation at short-distances, with a concomitant drastic 
reduction in the quasiparticle (QP) weight Z. In general, it appears that 
holes tend to induce one-dimensional-like spin arrangements to improve 
their mobility. Using ladders it is also shown that the 
very small J/t$\sim$0.1 regime of the standard t-J model may be more
realistic than anticipated in previous investigations, since such regime
shares several properties with those found in the extended model at
realistic couplings. 
Another goal of the present article is to provide
additional information on the recently discussed tendencies to 
stripe formation and spin incommensurability reported for the extended t-J model. 
These tendencies are here illustrated  with several examples.

\end{abstract}
\vskip2pc]
\narrowtext

\section{Introduction}

The study of high temperature superconductors continues
attracting the attention of the Condensed Matter community. In
recent years, much effort has been devoted to the understanding
of the spin incommensurability that appears in neutron scattering
experiments for some of these compounds\cite{tranquada}. 
Tranquada's interpretation
of the experiments is based on ``stripes'' where charge is confined
to one-dimensional (1D) paths in the crystal, with an average hole charge
n$_h$=0.5, namely one hole for every two sites along the 
stripe\cite{tranquada}. The stripe interpretation appears robust in the one-layer material
$\rm La_{2-x} Sr_x Cu O_4$, although it is still controversial in
the bilayer $\rm Y Ba_2 Cu_3 O_{6+\delta}$\cite{mook}. Experimental
results compatible with $metallic$ stripes
have also been reported using other techniques\cite{uchida}.
On the theory front, early studies discussed the presence of stripes 
in Hubbard and t-J models, although typically with density 
n$_h$=1.0\cite{zaanen,emery}. 
Recently, the existence of n$_h<$ 1 
stripes was discussed using computational techniques directly in the
t-J model at intermediate coupling J/t, without the need of long-range
Coulomb interactions\cite{white1}. 
Sometimes these stripes are described as a condensation
of d-wave pairs\cite{white2}. 
Stripes appear also in Monte Carlo
studies of the spin-fermion model\cite{moreo}. 
However, the origin of such complex spin-charge arrangement
is still under much discussion, and even its stabilization
remains controversial in the standard t-J model\cite{contro}.

In parallel with the developments in neutron scattering
experiments, in recent years 
photoemission techniques have provided a plethora of
information about the cuprates. In particular, the experimental
study of the
one-hole spectral function of the parent insulator
compound gave us useful information to judge the quality of
existing models for these materials\cite{arpes}. 
Based on the experimental one-hole dispersion
it has been convincingly shown that the t-J model is $not$ enough
to address the cuprates but extra terms must be added in order to
reproduce the photoemission results for the insulator\cite{eder}. These terms
appear in the form of extra hole hoppings, regulated in intensity by 
hopping amplitudes usually denoted by t$'$ and t$''$. The bare value
of these extra amplitudes is small compared with the nearest-neighbor
hopping t, but its influence is substantial since they link sites
belonging to the same sublattice and they are not
so heavily renormalized to smaller values as it occurs with t. The
origin of these extra hoppings was discussed before\cite{eder} and
they appear naturally in mappings from the three-band model for cuprates to a
one-band Hamiltonian, and also in band-structure calculations. In fact,
what is unnatural is to assume that the standard t-J model, 
with t$'$=t$''$=0.0, is valid for the cuprates without corrections. 
Unlike in gauge
theories, there are no renormalizability or symmetry arguments at work 
preventing the introduction of nonzero extra hoppings in models for
copper oxides. For
these reasons it is important to study the extended t-J model in detail
and, in particular, whether stripes occur in its ground state.

Recent studies of the extended t-J model have provided useful and
sometimes surprising information\cite{martins1,martins2}. 
For instance, it was observed that the one-hole QP
weight Z is considerably reduced by the addition of t$'$ and t$''$ at
fixed J/t. This result is particularly 
dramatic at momenta $(0,\pi)$ and $(\pi,0)$,
and its origin lies in the generation of across-the-hole
antiferromagnetic (AF) correlations, namely in the reference frame of the
mobile hole the two spins immediately next to it along one axis are
aligned $antiferromagnetically$\cite{martins1}. 
This tendency is the opposite as expected
from a vacancy in an antiferromagnet where those two spins should be 
ferromagnetically aligned, since they are in the same sublattice. Such a
curious result was also observed in previous DMRG studies of ladders and
planes using the standard t-J model by White and Scalapino\cite{white3}. 
Its importance and origin was recently discussed
by Martins et al.\cite{martins1} where its presence was
conjectured to be caused by tendencies in the system to
spin-charge separation at short distances. 
In other words, the kinetic energy of the
 hole with the extra mobility
induced by t$'$ and t$''$ dominates most of 
the physics of the one-hole problem and it tries to generate in its
vicinity an environment that allows for the hole to move without fighting against
the spin background. In one-dimension this is easy to set up, and indeed an 
antiferromagnetic across-the-hole environment is stabilized on chains
leading to spin-charge separation\cite{shiba}. 
In two-dimensions the hole tries to
create such an environment in its vicinity but it cannot
produce
a total spin-charge separation since frustration is induced when 
antiferromagnetic across-the-hole interactions are generated along
both axes. It is 
expected that for realistic values of the couplings, the spin-charge
separation will only be local 
(i.e. at short distances)\cite{martins1}. This result, obtained
using small square clusters\cite{martins1}, is clear
in the presence of nonzero t$'$ and t$''$ of the proper sign and
magnitude, but it also occurs in the standard t-J model in the regime
of very small J/t, which remains mostly unexplored. This interesting
 observation
redefines the relevant value of the parameter J/t from the number
usually used in most studies (0.3-0.4) to smaller values close to 0.1.

Also motivated by the results in Ref.\cite{martins1}, recently the
analysis of many holes at couplings that are expected to lead to robust
across-the-hole correlations was reported by Martins et
al.\cite{martins2}. Indications of stripes were obtained in this
case, similarly as those previously found in Ref.\cite{white1},
 rationalized as the natural tendency of locally spin-charge
separated states to avoid the spin frustration caused by the nontrivial 
spin environment that each hole generates. The hole density in these
stripes is in good agreement with experiments, and qualitatively the
results resemble those in the ``holons in a row'' picture of
Zaanen\cite{zaanen}. Tight 
hole bound-states are not needed to generate stripes.

These observations suggest that either the extended t-J model or
the standard t-J model at very
small coupling must be analyzed in detail if
the goal is to reproduce the physics of the striped regime of the cuprates.
In addition, it is important to confirm the 
generality of previous results\cite{martins1} 
by studying other systems with similar physics but where computational
studies can be carried out on larger clusters.
For this purpose, here 
$ladders$ are investigated using the extended and standard
t-J models in the coupling regime where in Ref.\cite{martins1} 
the robust across-the-hole feature was observed on small square
clusters. Our main result is that these robust correlations are
also very clearly identified on ladders, even with only two legs, and,
thus, its presence
is more general than naively anticipated. 
Based on our results it is clear that holes on ladders tend to form
one-dimensional (1D) like environments in their vicinity
 to improve their kinetic energy leading
to an effective reduction in dimensionality.
Spin incommensurability is generated by this procedure,
a novel result in two-leg ladder systems to the best of our knowledge, although
its presence in two dimensions has been known for some
time\cite{review}.

The present paper has a second goal which is the detailed analysis of
stripe formation upon doping of ladders with many holes. For this purpose
here the recent results of Ref.\cite{martins2} where stripes were
observed are extended to 
couplings and parameters not reported before, 
to illustrate and confirm the tendencies toward
stripe formation in this context. The proposed rationalization for this stripe
stabilization was already discussed\cite{martins2} and it is
based on the sharing of the locally spin-charge separated
spin environment created by
each individual hole, thus avoiding spin frustrating effects.

The model used here is the extended t-J model defined as
$$
\rm
H = J \sum_{\langle {\bf ij} \rangle} 
({{{\bf S}_{\bf i}}\cdot{{\bf S}_{\bf j}}}-{{1}\over{4}}n_{\bf i}n_{\bf j})
- \sum_{ {\bf im} } t_{\bf im} (c^\dagger_{\bf i} c_{\bf m} + h.c.),
\eqno(1)
$$
where $\rm t_{\bf im}$ is t for nearest-neighbors (NN), t$'$ for 
next NN, and t$''$ for next next NN sites, and
zero otherwise. The scale of energy will be t=1 unless otherwise
specified. The rest of the notation is standard. 
Comparison with PES experiments\cite{arpes} showed that 
t$'$=-0.35 and
t$''$=0.25 are relevant to explain PES data\cite{eder,martins1}. 
To simplify our studies the ratio t$'$/t$''$ will be fixed to -1.4 
in most of our analysis, although other ratios will be used in some cases. 
As numerical techniques, the 
Density Matrix Renormalization Group (DMRG)\cite{white1,error},
Exact Diagonalization\cite{review}, and 
an algorithm using
a small fraction of the ladder rung-basis 
(optimized reduced-basis approximation, or ORBA\cite{orba}) are here used. 
The paper is organized by the number of legs of the ladders considered,
from two to six. Analytical results are discussed after the numerical
methods, illustrating the appearance of tendencies toward across-the-hole
antiferromagnetism in two dimensional systems. In the last section it is
concluded that the mostly unexplored extended t-J model contains
interesting physics related to the cuprates, that deserves further work.

\section{Two-Leg Ladders:}

\subsection{ One Hole:}

The present computational analysis of ladder systems starts with the 
two-leg ladder case. Previous literature has shown that in the
absence of t$'$ and t$''$ hoppings, and at intermediate values of
J, such as 0.4, one hole behaves as expected from a carrier in
an antiferromagnet even if the ladder spin background   
has only short-range
magnetic order\cite{ladders}. In other words, a hole creates a spin distortion
in its vicinity (spin polaron) and the one-hole ground state
has a finite QP weight Z. Two of these spin polarons 
bind into a hole pair that leads to superconducting correlations
upon further doping\cite{ladders}. 
However, in this paper our goal is to analyze
a region of parameter space not studied before, to the best of
our knowledge,
where the hole is expected to be substantially more mobile. For
this purpose J is reduced to values between 0.1 and 0.2, still
within the domain considered realistic in studies of the cuprates,
and t$'$, t$''$ are made nonzero and of the sign and magnitude
as suggested by photoemission experiments. 
At such ``small'' values
of J/t, the spin background no longer is expected to fully control
the behavior of the hole but the spins have to arrange in such
a way that the hole kinetic energy is optimized (but still without leading
to a ferromagnetic state that may be the best configuration at
J/t$\sim$0.0).

Consider first Exact Diagonalization results in the one-hole sector.
In Fig.1a, the spin-spin correlations are shown for the case where
the hole is projected on the (arbitrary) 
site indicated, from the full one-hole 
lowest-energy state with momentum ${\bf k}$=$(\pi,0)$,
working at J=0.2, t$'$=-0.35 and t$''$=0.25. It can be observed 
that at a distance of about
three lattice spacings from the hole the spin correlations are
similar to the ones expected for an undoped ladder, with robust antiferromagnetism
along both the rungs and legs. 
However, in the vicinity of the hole the spin
correlations are substantially altered. In this case 
the spins belonging to the
same rungs are no longer strongly antiferromagnetically
correlated and the system in the vicinity of the
hole resembles a pair of weakly-coupled portions of a chain, one per
leg. In
this respect it appears that an ``effective'' transition from a two-leg
ladder to 1D chains has occurred near the hole, a surprising result.
Holes seem to optimize their energy by creating 1D environments in its
vicinity. Note the presence of robust across-the-hole spin correlation on
the upper leg of Fig.1a, which also appears in the exact solution
of the 1D Hubbard model due to spin-charge separation. Fig.1a suggests
that the spin associated to the hole is spread uniformly in the
1D-like segment dynamically generated, and in this respect spin-charge
separation occurs locally, as anticipated in the Introduction and as
it was reported in Ref.\cite{martins1} using small square clusters.
Note in Fig.1a that the spin belonging to the same rung as the hole
appears as ``free'' and there is a strong AF bond at distance of two
lattice spacings across it, as it occurs across-the-hole.

\begin{figure}
\begin {center}
\mbox{\psfig{figure=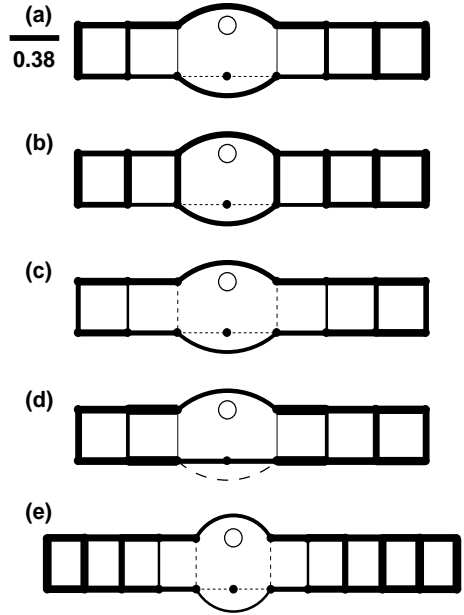,height=3.2in,width=2.4in}}
\end{center}
\caption{
 Spin-spin correlations in the state with the lowest energy
 of one hole with momentum ${\bf k}$, at the 
 couplings described below.
 The data were obtained on the two-leg ladder using the
 Lanczos method on 2$\times$8 (a-d) and 2$\times$12 (e) clusters,
 with periodic boundary conditions (PBC) along the leg direction. 
 Shown are results when the hole is projected 
 from the state under consideration at the
 site indicated by an open circle. 
 The dark lines between sites represent
 the absolute value of the (antiferromagnetic) 
 $\langle$${{{\bf S}_{\bf i}}\cdot{{\bf S}_{\bf j}}}$$\rangle$
 spin correlation 
 between the spins located at those sites (scale shown in (a)). 
 Dashed lines
 indicate weak ferromagnetic correlations. 
 Results at distance
 of two lattice spacings are shown only near the hole.
 Note the presence of strong 
 across-the-hole antiferromagnetic bonds in all the cases.
 (a) corresponds to J=0.2, t$'$=-0.35, t$''$=0.25 and
 $\bf k$=($\pi$,0). Far from the hole the spin correlations
 are as in undoped two-leg ladders, while near the hole
 they are much distorted, and they are robust mainly along the legs.
 (b) is the same as (a) but for
 t$'$=t$''$=0.0. (c) Results at J=0.1, t$'$=-0.35, t$''$=0.25 and
 $\bf k$=($\pi$,0). Note the appearance of some weak
 ferromagnetic rungs at this small value of J. (d) Same as (a)
 but for $\bf k$=($\pi/2$,0). (e) corresponds to J=0.2, 
 t$'$=-0.35, t$''$=0.25 and $\bf k$=($\pi$,$\pi$).
}
\end{figure}

The comparison of Figs.1a and 1b, the latter obtained without extra
hoppings, 
shows that t$'$ and t$''$ are important at J=0.2 to regulate
the size of the 1D environment around the hole: as t$'$, t$''$ grow
in magnitude, the size of the 1D-like region also grows. This same effect
is produced reducing J at fixed t$'$,t$''$ as shown in Fig.1c: here
the distortion is larger than at J=0.2 and even weak ferromagnetic
links are generated between the chains, an unexpected result quite
different from the well-established physics of undoped ladders at
intermediate J.
The results here are also weakly dependent on the momentum: in Fig.1d,
the case of ${\bf k}$=$(\pi/2,0)$ (momentum of the overall
ground-state of one hole) 
presents a distortion similar in size as at 
${\bf k}$=$(\pi,0)$, although the AF long-bond in the leg opposite to where
the hole 
is projected (Fig.1a) is no longer present. However, the long-bond across-
the-hole is still robust.
Fig.1e is another case illustrating the
momentum dependence in the problem, corresponding this time
to ${\bf k}$=$(\pi,\pi)$. Overall, it can be concluded that
the lowest one-hole energy states for all momenta studied here, 
and in large regions of parameter space, present
very similar spin arrangements.
The across-the-hole feature emphasized in Ref.\cite{martins1}
is very robust in all cases
and in this sense there is a clear $\pi$-shift across-
the-hole even in the simple case of a two-leg ladder, similarly
as it occurs in 2D systems. The dynamically
induced 1D-like regions near the carrier are also clear in our studies.

\vspace{-0.2in}
\begin{figure}
\begin {center}
\mbox{\psfig{figure=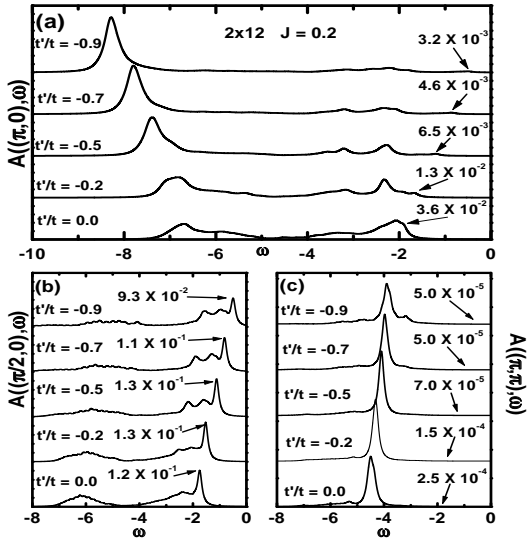,height=3.5in,width=3.2in}}
\end{center}
\vspace{-0.3in}
\caption{One-hole spectral function $\rm A({\bf k},\omega)$ obtained
 exactly from the undoped ground state of a 2$\times$12 
 cluster working at J=0.2, and t$'$=-1.4t$''$ fixed,
 parametric with t$'$. The arrows indicate the position of 
 the first pole in the spectra and the numbers next to them are the
 weights Z (normalized such
 that 0$\leq$Z$\leq$1). (a) corresponds to $\bf k$=($\pi$,0),
 (b) to $\bf k$=($\pi$/2,0), and (c) to $\bf k$=$(\pi,\pi)$.}
\end{figure}

 The generation of a quasi-1D structure in the vicinity of the
hole and the expected spread of the hole's spin-1/2 over
several lattice spacings (spin-charge separation at short distances)
causes a drastic reduction in the QP weight Z.
In Fig.2, the one-hole spectral function A(${\bf k}$,$\omega$) is presented
at three different momenta. The results are shown as it is usual
in experimental photoemission literature with the lower
energy states appearing near $\omega$=0, reference energy which
is located at an arbitrary position in this half-filled case,
and the rest of the states running to the left. 
Data for J=0.2 and several values of
t$'$, at fixed t$'$/t$''$=-1.4, are presented, together with the weight Z
and position of the first state (arrows). 
Fig.2a shows that Z at this value of J
and ${\bf k}$=$(\pi,0)$ is already small even in the absence 
of extra hoppings, result similar to those
observed in previous investigations of the t-J model on square
lattices\cite{review}. However, it is
remarkable the rapid reduction of Z with increasing $|$t$'$$|$, to
values that in practice are basically negligible by the time
these extra hoppings reach its realistic values. This is correlated
with the appearance of the complex quasi-1D structure discussed in
Fig.1 \cite{comm1} 
A similar phenomenon but even more pronounced exists for
${\bf k}$=$(\pi,\pi)$ (Fig.2c), where Z is negligible even without
extra hoppings. On the other hand, the results at
${\bf k}$=$(\pi/2,0)$ (Fig.2b), while still corresponding to small
values of Z of the order of 10\% of the maximum, do not show such
a dramatic reduction with increasing t$'$. In this case the spin
appears still located in the vicinity of the hole, although it is
spread over several sites. Z becomes negligible at this momentum
only at values of $|$t$'$$|$ larger than shown in the figure.

\vspace{-0.2in}
\begin{figure}
\begin {center}
\mbox{\psfig{figure=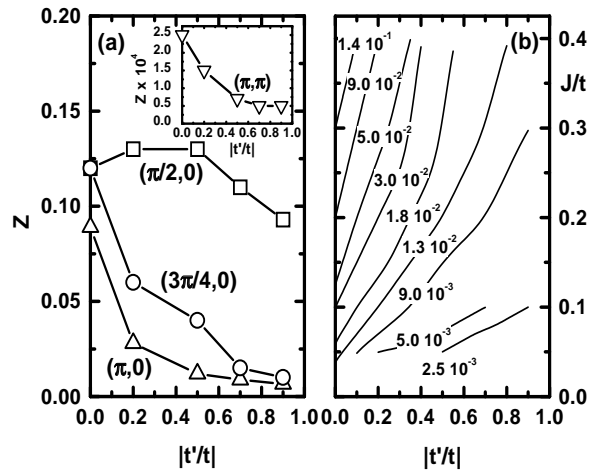,height=4.5in,width=3.2in}}
\end{center}
\vspace{-1.8in}
\caption{(a) Weight Z of the lowest energy state in the $\rm A({\bf k},\omega)$
 one-hole spectral function of exactly solved 2$\times$12 clusters 
 at J=0.2, t$'$=-1.4t$''$, as a function of $|$t$'$/t$|$.
 Z is normalized such that 0$\leq$Z$\leq$1. 
 Shown are results at the momenta indicated. In the inset the
 same information is given but for the nearly negligible lowest-pole
 weight of $\bf k$=$(\pi,\pi)$. 
 (b) Lines of constant Z weight (value indicated) 
 in the plane J/t--$|$t$'$/t$|$, with
 t$'$=-1.4t$''$ fixed, corresponding to the lowest-energy
 pole in the $\bf k$=$(\pi,0)$ subspace. 
 Note the drastic reduction in Z as J is reduced from 0.4
 and/or $|$t$'$$|$ is increased from 0.0. The results are 
 interpolations using a fine grid of points obtained 
 from the 2$\times$8 cluster solved exactly.
}
\end{figure}

The results for the QP weight Z are more explicitly illustrated
in Fig.3a where its $|$t$'$$|$ dependence is shown for several momenta at
fixed J=0.2. Clearly for all momenta the presence of t$'$ eventually
leads to a drastic reduction of the Z value, although for ${\bf
k}$=$(\pi,0)$ it happens rapidly with increasing t$'$
while at ${\bf k}$=$(\pi/2,0)$ it needs
a t$'$ larger than the nearest-neighbors hopping t.
In Fig.3b the lines of constant weight Z corresponding to
${\bf k}$=$(\pi,0)$ are shown.  It is clear that Z at this momentum
rapidly reduces its value, both at fixed J increasing $|$t$'$$|$ or
for fixed hoppings decreasing J. 
Although a careful finite size scaling is difficult, 
previous experience with two-dimensional clusters
suggest that Z tends to decrease as the lattice size grows. As a 
consequence it is expected that the values reported in Figs.2  and 3
will actually be even smaller for a very long two-leg ladder.
Note also that the lines of constant
Z suggest that the physics of, say, intermediate J and finite t$'$
may be smoothly connected to that of small J and zero t$'$,  quite
similarly as it happens on square clusters\cite{martins1}. Then,
it may occur that analyzing the regime of abnormally
small J ($\sim$0.1) of
the t-J model may effectively account for the presence of extra
hoppings, as was conjectured in Ref.\cite{martins1}. This assumption
is interesting due to the anticipated 
unusual properties of the small J/t region
of the t-J model, and it avoids the somewhat aesthetically unpleasant
use of extra hoppings in the present theoretical studies.

\subsection{ Several holes:}

The previous subsection showed that the two-leg ladder system presents
exotic behavior in the one-hole sector when the regime of small J
is investigated or alternatively when at a fixed J the hoppings
t$'$-t$''$ are switched on. It is important to explore what occurs
for more holes in the problem, and Fig.4 contains exact results for
the case of two holes on an 2$\times$8 ladder. Data is shown for the
case when the two holes are projected to its most probable location
in the ground-state,
which in the cases studied coincide with the maximum distance among them
allowed in the cluster. The spin structures in Fig.4 near the holes clearly
resemble those found around individual holes in the previous subsection.
In particular the across-the-hole AF correlation is very prominent
in all cases shown in Fig.4.
This structure naturally leads to spin incommensurability, 
as observed in the spin structure factor presented in Fig.5a\cite{comm2}.
It is interesting to observe that the coupling between chains can be
weakly $ferromagnetic$ in some cases (Fig.4a), weakly AF in others
(Fig.4b), or even virtually negligible (Fig.4c) depending on the
actual couplings used. Overall it is clear that
the legs are approximately decorrelated upon hole doping
at the couplings analyzed here, and each leg behaves 
as a 1D-chain described by the same original t-J model. Each individual
leg resembles the expected result for the 1D Hubbard model at
large U\cite{shiba}, as discussed before for 2D clusters\cite{martins1}. 
It is remarkable that very recently nuclear spin relaxation results for 
two-leg ladder materials have also reported the decoupling of the two legs 
at high temperature\cite{fujiyama}. The analysis of the relation between our 
results and those of Ref.\cite{fujiyama} deserves further work.

\begin{figure}
\begin {center}
\mbox{\psfig{figure=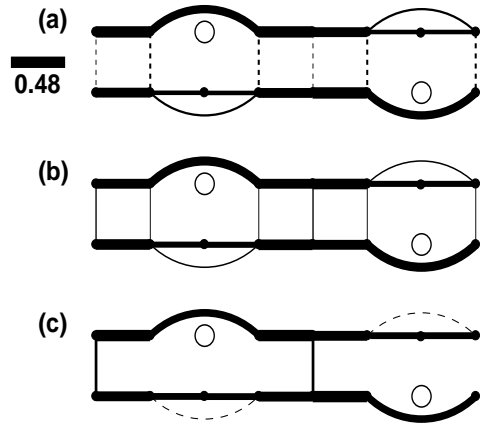,height=2.2in,width=2.5in}}
\end{center}
\caption{Spin-spin correlations in the state with the lowest energy
 of two holes, at the couplings described below.
 The data were obtained exactly on the two-leg ladder using the
 Lanczos method on 2$\times$8 clusters with PBC along the legs. 
 Shown are results when the holes
 are projected from the state under consideration at the
 sites indicated by open circles. These hole positions are
 the most probable in the two-holes ground-state.
 The dark lines between sites represent
 the strength of the antiferromagnetic spin correlation 
 between the spins located at those sites (scale shown in (a)). Dashed lines
 indicate weak ferromagnetic correlations, and the
 missing vertical correlations indicate links with virtually
 negligible antiferromagnetic spin correlations.
 Results at distance
 of two lattice spacings are shown only near the hole.
 Note the presence of strong 
 across-the-hole antiferromagnetic bonds in all the cases, as in the
 one-hole problem.
 Note also the clear tendency of spins in opposite legs to be nearly
 decorrelated.
 (a) corresponds to J=0.2, t$'$=-0.35, and t$''$=0.25.
 (b) corresponds to J=0.2, and t$'$=t$''$=0.0.
 (c) corresponds to J=0.4, t$'$=-0.35, and t$''$=0.25.}
\end{figure}

In the two-hole ground-state the expectation value of the number
operator for a particular momentum is shown in Fig.5b. Clearly the
momenta $\rm {k_x}$ dominating in this state 
is not only $\pi/2$, the momentum of the
lowest energy state of the one-hole sector, but actually $3\pi/4$
and $\pi$ are also important as well as the $\rm {k_y}$=$\pi$ sector.
Then, there is no indication of ``hole-pockets''
in the doped system, and momenta over a wide range contribute to
the two-hole ground-state. This suggests that the one-hole
states with the largest weight in the  two-hole ground-state have
the across-the-hole structure and local spin-charge separation.

\begin{figure}
\begin {center}
\mbox{\psfig{figure=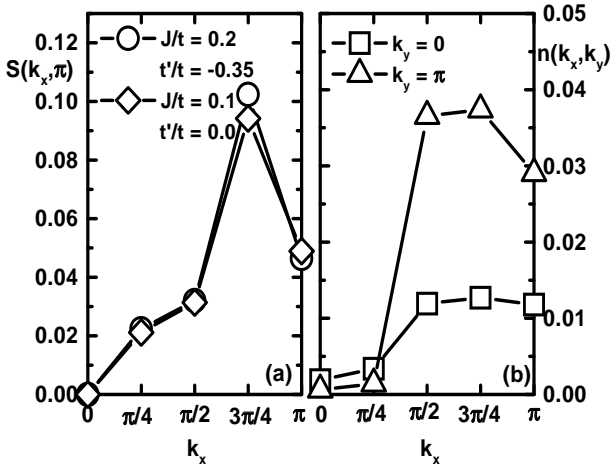,height=4.5in,width=3.2in}}
\end{center}
\vspace{-1.8in}
\caption{(a) Spin structure factor S($\rm k_x$,$\rm k_y$=$\pi$) vs $\rm k_x$
 for the couplings J and t$'$ indicated (with t$'$=-1.4t$''$ fixed),
 using the two holes ground-state of the 2$\times$8 cluster.
 Clear tendencies toward incommensurate 
correlations are observed in our study, as
 exemplified here,
 either at small J without t$'$-t$''$ hoppings, or increasing the values
 of the latter at fixed J.
 (b) Exact n($\rm k_x$,$\rm k_y$)=$\sum_{\sigma}$$\langle$ 
 $c^{\dagger}_{{\bf k}\sigma} c_{{\bf k}\sigma}$$\rangle$ vs $\rm k_x$
 for the $\rm k_y$ momenta indicated, obtained using the ground-state
 of the two holes 2$\times$8 cluster and $c_{{\bf k}\sigma}=
 \frac{1}{\sqrt{N}}\sum_{\bf j}e^{-i{\bf k}{\bf j}}c_{{\bf j}\sigma }$, 
 where N is the number of sites in the cluster, ${\bf j}$ is a site index 
 and $\sigma =+,-$. The
 couplings are J=0.2, t$'$=-0.35 and t$''$=0.25.
}
\end{figure}

The charge structure factor was also calculated in these investigations.
In Figs.6a-b exact results on a 2$\times$8 cluster and DMRG results
on 2$\times$16 and 2$\times$32 clusters, all at hole
density x=0.125, are shown. They are very
similar, illustrating the lack of strong size effects systematically 
observed in our studies of two-leg ladders. 
The results obtained in the more traditionally studied case of
J=0.4, t$'$=t$''$=0.0 are also shown in Figs.6c-d. They are
qualitatively different from those in Figs.6a-b, difference
likely related with the formation of
the quasi-1D regions mentioned before.

\begin{figure}
\begin {center}
\mbox{\psfig{figure=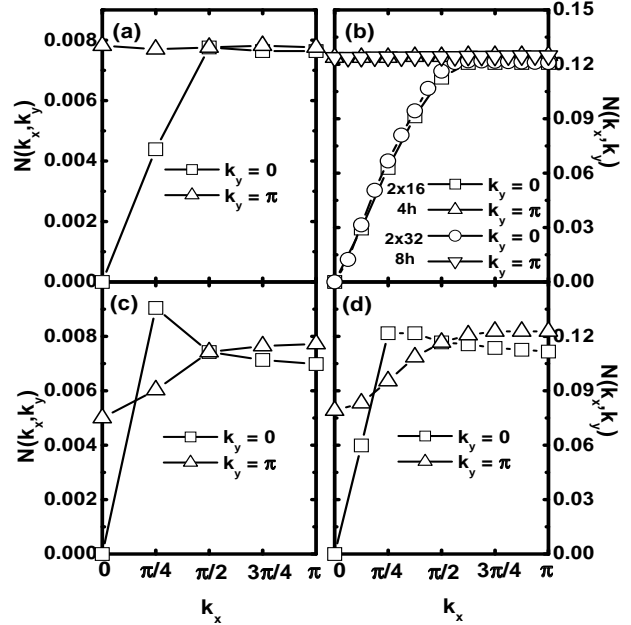,height=3.8in,width=3.2in}}
\end{center}
\vspace{-0.3in}
\caption{Charge structure factor N($\rm k_x$,$\rm k_y$) vs $\rm k_x$, 
 parametric with $\rm k_y$ as indicated. 
 (a) corresponds to J=0.2, t$'$=-0.35, and t$''$=0.25 using the
 two holes ground-state of the 2$\times$8 cluster solved exactly.
 (b) Same as (a) but using 2$\times$16 and 2$\times$32
 clusters and 4 and 8 holes, respectively, studied with 
 the DMRG technique. 
 (c) Same as (a) but for J=0.4, and t$'$=t$''$=0.0.
 (d) Same as (c) but on a 2$\times$16 cluster with 4 holes
 studied with the DMRG technique.
 Comparing (a) with (b), and (c) with (d), it appears that 
 finite size effects are small.}
\end{figure}


\section{Three-Leg Ladders:}

\subsection{One hole:}

Results for three-leg ladders have also been
gathered in this work. In Fig.7 the hole is projected at the location shown,
from the lowest-energy state at the momenta indicated in the caption.
In Fig.7a results are presented at J=0.1 and nonzero t$'$-t$''$.
The formation of the across-the-hole structure is very clear, in
both directions. Along the vertical one, a strong AF bond is formed
between the two spins of the rung where the hole is located. Its
strength is close to that of a perfect singlet and, thus, the coupling
of those spins with the rest is small. This structure is present
at all the couplings investigated as shown in Figs.7a-c, even 
intermediate and large J, and it is
also clearly present in the exact solution of a three-site t-J model
at many couplings. In the small J/t regime of the standard t-J model, 
and in the extended t-J model as well, this strong AF bond denotes a precursor of the 
$\pi$-shift across the stripe that will form (along the PBC direction) 
as the hole-density increases. 
Regarding the horizontal AF bond (along the legs), its strength
depends on the particular value of
J and t$'$-t$''$. In the intermediate coupling
region often studied, exemplified by J=0.5 and no extra hoppings as
in Fig.7c, there is no across-the-hole AF correlation along the legs.
On the other hand, for the couplings of Fig.7a, the clear 1D-like
AF segment along the central leg induced in the vicinity of the 
hole suggests
a larger mobility of the carrier and  generation of 1D-like segments, 
as discussed in the Introduction and for two-leg ladders.

\begin{figure}
\begin {center}
\mbox{\psfig{figure=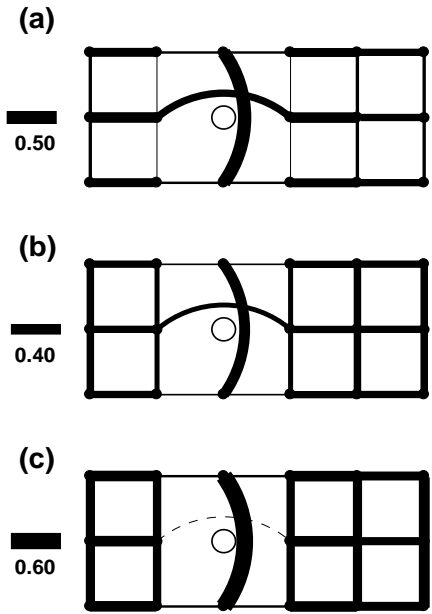,height=3.2in,width=2.3in}}
\end{center}
\caption{Spin-spin correlations in the state with the lowest energy
 of one hole with momentum ${\bf k}$, 
 at the couplings described below.
 The data were obtained exactly on the three-leg ladder using the
 Lanczos method on 3$\times$6 clusters. PBC are used in the long
 direction and open boundary conditions (OBC) in the short one. 
 Shown are results when the hole is projected at the
 site indicated by an open circle from the state under consideration.
 The dark lines between sites represent
 the absolute value of the (antiferromagnetic) spin correlation 
 between the spins located at those sites (scale shown). 
 The dashed line in (c)
 indicates a weak ferromagnetic correlation. 
 Results at distance
 of two lattice spacings are shown only near the hole.
 (a) corresponds to ${\bf k}$=($\pi$/3,0), J=0.1, t$'$=-0.35, and t$''$=0.25.
 (b) corresponds to ${\bf k}$=($\pi$/3,$\pi$), J=0.1, and t$'$=t$''$=0.0.
 (c) corresponds to ${\bf k}$=($\pi$/3,$\pi$), J=0.5, and t$'$=t$''$=0.0.}
\end{figure}

The tendency toward local spin-charge separation can be studied 
further if the mean-value of the spin in the z-direction
is calculated when the hole is projected from its ground-state
to a given site, as done before in Figs.7a-c. Since removing one spin
from a cluster with an even number of sites
creates states with spin-1/2, together with the hole in Fig.8
there must be a spin-1/2 spread over the cluster.
The results are
shown in Fig.8a for the case of small J and nonzero t$'$-t$''$.
In this situation it is clear that the z-component of the spin 1/2 is distributed approximately
uniformly along the central leg where the hole is projected (the
spin in the outer legs is negligible). In
agreement with the introductory discussion, the mobile hole creates
a 1D environment (along the central leg in this case)
to help in its propagation, and in this context the
spin and charge are separated.
On the other hand, the results corresponding to an intermediate
J of value 0.5 and no extra hoppings corresponds to a staggered 
spin background surrounding the hole (Fig.8b), similar to results for
a vacancy in a N\'eel background, with the exception of the
two spins in the same rung as the hole, which try to form a strong
singlet and thus its mean z-component spin tends to vanish.
There is a clear {\it qualitative} difference between Figs.8a and b,
caused by the high mobility of the hole in the presence of a small
J and nonzero hoppings beyond nearest-neighbors. 
The QP weights Z (not shown) associated to the structure of Fig.8a
tend to vanish at ${\bf k}$=$(0,\pi)$ and others, as found before for
two-leg clusters and small two-dimensional systems.

\begin{figure}
\begin {center}
\mbox{\psfig{figure=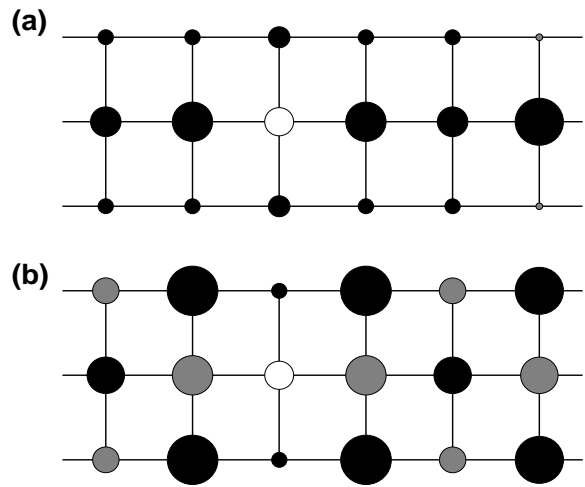,height=2.5in,width=3.0in,angle=-90}}
\end{center}
\caption{Mean value of the z-component of the spin ($\langle S^z_{\bf i}
\rangle$)  
 denoted by the area of the circles, obtained from 
 the one-hole ground-state of the 3$\times$6 cluster with the hole
 projected at the open circle position. Black (gray) circles indicate
 positive (negative) values.
 (a) corresponds to J=0.1, t$'$=-0.35, and t$''$=0.25, while
 (b) corresponds to J=0.5, t$'$=t$''$=0.0.}
\end{figure}

\subsection{Many holes:}

When two holes are considered on the three-leg ladder the situation
is similar as in the case already described
of two-legs, namely each individual hole
carries a spin arrangement in its vicinity similar to that found in the
one hole case. Results are shown in Fig.9a at small J and Fig.9b
at intermediate J, in both cases without t$'$-t$''$. In the first
case, AF bonds across-the-hole are found in both directions, while
in the second the bond along the legs turns ferromagnetic as for
vacancies in a N\'eel background and as in Fig.7c for one mobile
hole\cite{comm4}. 

\begin{figure}
\begin {center}
\mbox{\psfig{figure=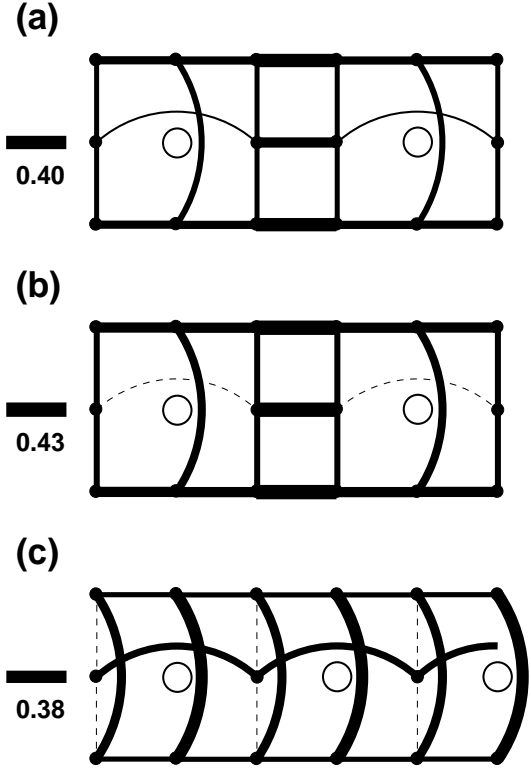,height=4.0in,width=2.8in}}
\end{center}
\caption{Same convention 
as in Fig.7 but now for the two holes ground-state working
 at: (a) J=0.2, and t$'$=t$''$=0.0, and (b) J=0.5, and t$'$=t$''$=0.0.
In this case the most likely configuration is a tight bound 
state (not shown). (c) Same as (a) but for the three holes ground-state 
at J=0.25, t$'$=-0.35, t$''$=0.25. In all cases the momentum is (0,0).}
\end{figure}

If three holes are considered at small J (Fig.9c) the most likely 
hole configuration corresponds to a n$_h$=0.5 stripe along the central
leg, with clear and robust AF bonds across it joining the two outer
legs (in spite of having open boundary conditions along the rungs).
Within the central leg or stripe, AF bonds are formed across-the-hole,
as in a 1D system with the same Hamiltonian. Fig.9c qualitatively
indicates the dynamical separation of a chain subsystem carrying the
charge from the rest of the spins which are correlated as if that chain
would not exist. As emphasized in other parts of this paper and in
Refs.\cite{martins1,martins2}, this
appears to be the way in which the system achieves
a partial separation of spin and charge in two dimensions.

\section{Four- and Six-leg Ladders:}

\subsection{One Hole:}

Results similar to those found in the case of two- and three-leg ladders
are also observed with four legs. Results for the case of one hole
with momentum ${\bf k}$=$(\pi,0)$ are shown in Fig.10 on
a 4$\times$6 cluster solved exactly using the Lanczos method. 
In Fig.10a results for small J and nonzero t$'$-t$''$ are presented: 
here the AF bonds across-the-hole can be observed and 1D-like
segments are created near the hole in both directions\cite{comm5}. 
The effect is amplified if at fixed J the values of the hoppings t$'$-t$''$
are increased. Fig.10b contains the results using abnormally large
extra hoppings: now the AF bonds across-the-hole are quite robust,
with a strength that grows as the extra hoppings grow
in magnitude. There is a smooth connection between the physically
acceptable values of t$'$-t$''$ and those used in Fig.10b to amplify the
effect. Such a connection suggests a common origin to the structure.

\begin{figure}
\begin {center}
\mbox{\psfig{figure=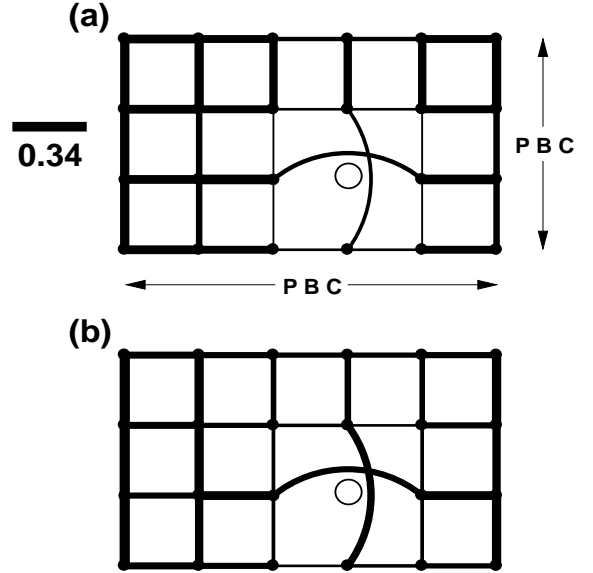,height=3.0in,width=3.0in}}
\end{center}
\caption{Spin-spin correlations in the state with the lowest energy
 of one-hole with momentum ${\bf k}$, at the 
 couplings described below for a four-leg ladder.
 The data were obtained exactly using the
 Lanczos method on a 4$\times$6 cluster with PBC in both directions. 
 Shown are results when the hole is projected at the
 site indicated by an open circle from the state under consideration.
 The dark lines between sites represent
 the strength of the antiferromagnetic spin correlation 
 between the spins located at those sites (scale shown in (a)). 
 Results at distance
 of two lattice spacings are shown only near the hole.
 (a) corresponds to J=0.2, t$'$=-0.35, t$''$=0.25 and
 $\bf k$=($\pi$,0). Far from the hole the spin correlations
 are as in undoped ladders, while near the hole
 they are much distorted. Note the clear presence of the across-the-hole
 antiferromagnetic bonds emphasized in this work.
 (b) Same as (a) but for t$'$=-1.50, t$''$=1.07 (keeping the same
 ratio t$'$/t$''$ as in (a)). These large values of the extra hoppings
 are used to increase the magnitude of the effect.}
\end{figure}

The one-hole spectral function $A({\bf k},\omega)$ is shown in Figs.11a-b
for the momenta indicated and as a function of $|$t$'$$|$. At 
$(0,\pi)$ the QP weight Z rapidly decreases with an
increasing t$'$ hopping amplitude, while at a momenta closer to the
expected ground-state momentum of one hole it remains more robust,
but it eventually tends to vanish at large $|$t$'$$|$ (see also
Fig.12). These results are very similar to those observed for two-leg
ladders, for planes\cite{martins1},
and also for three-leg ladders (not shown).

\vspace{-0.1in}
\begin{figure}
\begin {center}
\mbox{\psfig{figure=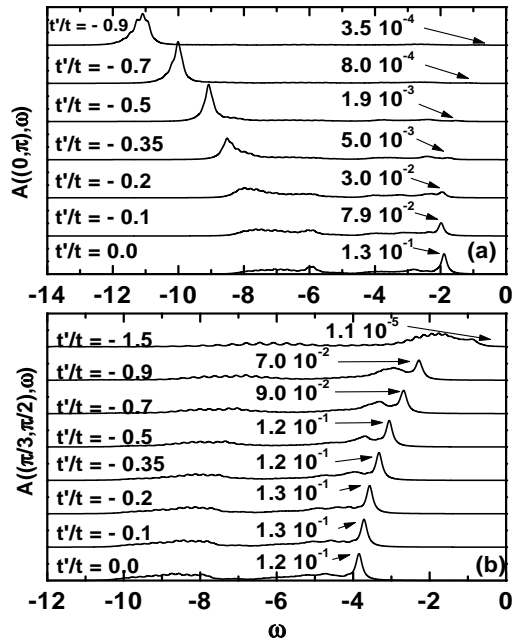,height=4.0in,width=3.2in}}
\end{center}
\vspace{-0.4in}
\caption{One-hole spectral function $\rm A({\bf k},\omega)$ obtained
 exactly from the undoped ground state of the 4$\times$6 
 cluster with PBC in both directions, working at J=0.2, with t$'$=-1.4t$''$ fixed,
 and parametric with t$'$. The arrows indicate the position of 
 the first pole in the spectra and its weight Z (normalized such
 that 0$\leq$Z$\leq$1). (a) corresponds to $\bf k$=(0,$\pi$) and
 (b) to $\bf k$=($\pi/3$,$\pi/2$) 
 (the closest to $\bf k$=($\pi/2$,$\pi/2$) in the cluster 
 considered here).}
\end{figure}

\vspace{-0.1in}
\begin{figure}
\begin {center}
\mbox{\psfig{figure=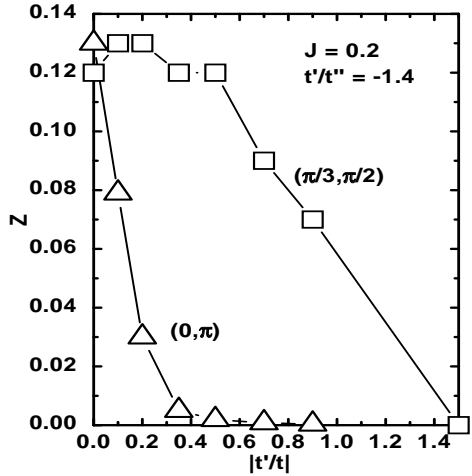,height=3.0in,width=2.8in}}
\end{center}
\vspace{-0.4in}
\caption{Weight Z of the lowest energy state in the $\rm A({\bf k},\omega)$
 one-hole spectral function of exactly solved 4$\times$6 clusters 
 at J=0.2, t$'$=-1.4t$''$, as a function of $|$t$'$/t$|$.
 Z is normalized such that 0$\leq$Z$\leq$1. 
 Shown are results at the momenta indicated.}
\end{figure}

\subsection{Many Holes:}

To clarify the behavior of two holes on four-leg ladders, 
consider the use of cylindrical boundary conditions (OBC in
one direction and PBC in the other\cite{white1}). Suppose the two-hole
ground-state is considered and one hole is projected to
a site belonging to the central legs, where the density of holes
is the largest. In this situation it is possible to
study the probability of finding
the other hole, with results shown in Fig.13a: the second hole
clearly prefers to be along the PBC direction and still within
the two central legs. Actually the largest probability is found
at a distance of two lattice spacings along the same leg.
The results can be interpreted as the formation of a loop of
charge which wraps around the direction with PBC, as discussed
recently in Ref.\cite{martins2} and also in Ref.\cite{white1}. The state
appears to correspond to a n$_h$=0.5 stripe, not rigid but
fluctuating in the direction perpendicular to it.

\begin{figure}
\begin {center}
\mbox{\psfig{figure=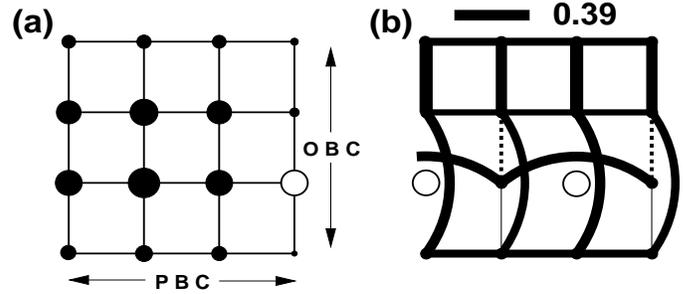,height=1.5in,width=3.5in}}
\end{center}
\caption{Results corresponding to the exact two holes ground-state of a
4$\times$4
 cluster with PBC in one direction and OBC in the other, 
 as indicated. The couplings are J=0.2, t$'$=-0.35, and t$''$=0.25.
 (a) corresponds to the case where one hole is projected from the
 ground-state into the location denoted by the open circle. The full
 circles have an area proportional to the probability of finding 
 the second hole at a given site. A loop of mobile holes is dynamically formed around
 the PBC direction.
 (b) AF spin-spin correlations with the convention
 followed in previous figures, for the case where the two holes are projected
 from the ground-state at the location indicated by open circles. This
 hole configuration is the one with the largest weight in the ground-state.
 Results at distance
 of two lattice spacings are shown only near the stripe.}
\end{figure}

In Fig.13b the spin correlations are shown for the holes
projected into the configuration
with the largest weight in the two-hole ground-state.
In agreement with the discussion associated with Fig.13a,
the holes appear to be forming a n$_h$=0.5 stripe
along the direction with PBC. Moreover, the spin correlations
across-the-stripe are clearly antiferromagnetic as in the
three-leg ladder (Fig.9c) and in 
experiments\cite{tranquada}, and the two spins of the four-site stripe 
considered here are correlated also antiferromagnetically.
This result is representative of a large number of
similar data gathered in our present analysis, namely it appears that in
general holes
tend to form loops of charge around the closed direction
if a system has cylindrical boundary conditions. This is
also in excellent agreement with the conclusions of earlier
work by White and Scalapino\cite{white1},
although their description of stripes is sometimes based on the
condensation of d-wave pairs\cite{white2} while ours is based on
a kinetic energy optimization (namely a one-hole problem).

The appearance of stripes based on the 4$\times$4 results of
Fig.13 needs to be confirmed increasing the cluster size. This
analysis was done in part in Ref.\cite{martins2} as discussed in
the Introduction, but here those
results are expanded and more details are provided.
Note that our present analysis using the DMRG method is restricted to the
t-J-t$'$ model, namely the t$''$ hopping will be considered to be
zero. The reason is that in the implementation of the
DMRG technique sites are
aligned along a one-dimensional pattern even for ladder
geometries, and a t$''$ hopping would link sites along this
equivalent chain which are located several lattice spacings from each other,
reducing the accuracy of the method. Nevertheless even without
t$''$ the stripes reported in Fig.13 and Ref.\cite{martins2} 
can be clearly observed on larger systems.
For example, consider in 
Fig.14 two holes on a 4$\times$12 cluster
with cylindrical boundary conditions (PBC along the rungs), J=0.5,
and t$'$=-0.3 (t$''$=0.0). 
One of the holes is projected to an arbitrary site
of one of the central rungs, where the hole density is the largest.
The distribution of the second hole around the projected one is
quite similar to the result observed on the 4$\times$4 cluster,
namely the largest chances are at distance of two lattice spacings
along the rung. This state resembles a bound-state in the sense that
the two holes are close to each other, but its shape is better
described as a stripe configuration or a loop of charge that wraps
around the short direction.

\begin{figure}
\begin {center}
\mbox{\psfig{figure=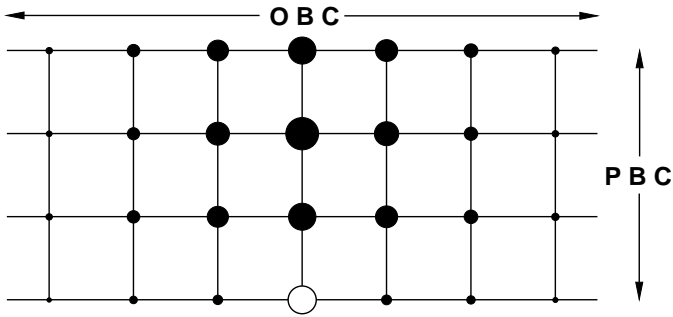,height=1.6in,width=3.5in}}
\end{center}
\caption{Results corresponding to the two holes ground-state studied with
 the DMRG technique on a 4$\times$12
 cluster with PBC in the short direction and OBC in the other, as indicated.
 The couplings are J=0.5, t$'$=-0.3, and t$''$=0.0. One of the holes
 is projected at the open circle site. The full circles denote the 
 probability of finding the second hole at a given site. A loop
 of mobile holes is formed wrapping around the rung direction.}
\end{figure}

If more holes are added to the four-leg ladder, it appears that
stripes similar to those observed in Fig.14 are formed. For example,
consider the case of four holes on a 4$\times$8 cluster at
small J and without t$'$ and t$''$. In Fig.15 the density of holes
is shown for the case where one hole is projected into one of the 
rungs with the largest density. It is clear from the figure that
in the vicinity of the projected hole there is a structure similar
to that observed for the two-hole ground state, namely a local
maximum in the hole density is observed at two lattice spacings from
the projected hole. In addition, clearly 
a large accumulation of charge appears on the other side of the
cluster from where the projected hole is located. This other sector populated
by holes involves two rungs in width, and it is framed in Fig.15.
Note that the density in that second stripe is very uniform showing
that there are no charge correlations between the two stripes (the two
charged loops are independent of each other).

\begin{figure}
\begin {center}
\mbox{\psfig{figure=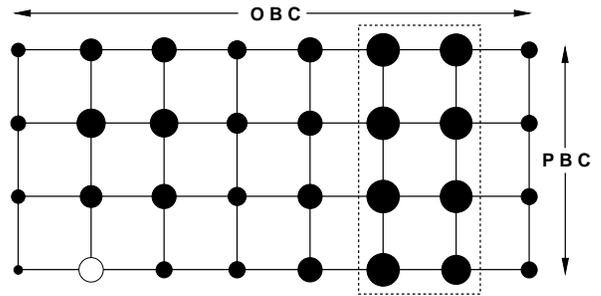,height=1.5in,width=3.0in}}
\end{center}
\caption{Results corresponding to the four holes ground-state treated with
 the DMRG technique on a 4$\times$8
 cluster with PBC in the short direction and OBC in the other, as indicated.
 The couplings are J=0.2, and t$'$=t$''$=0.0. One of the holes
 is projected at the open circle site. The full circles denote the 
 probability of finding another hole at a given site. Two loops
 of mobile holes appear to be formed wrapping around the PBC direction (one 
 of the two is framed).
}
\end{figure}

In Fig.16a the charge structure factor is shown for the case
of an 4$\times$8 cluster with four holes, studied with the DMRG
method at small J. The result resembles those obtained 
for two-leg ladders in Fig.6. In Fig.16b
the spin structure factor is shown for the same cluster and density.
Spin incommensurability clearly appears in this system, as remarked
before in Ref.\cite{martins2} 
working at other couplings, and as presented for the two-leg ladder
(Fig.5a) as well. It is clear that the two-, three- and four-leg ladder
systems share very similar physics in the charge and spin sectors.

\begin{figure}
\begin {center}
\mbox{\psfig{figure=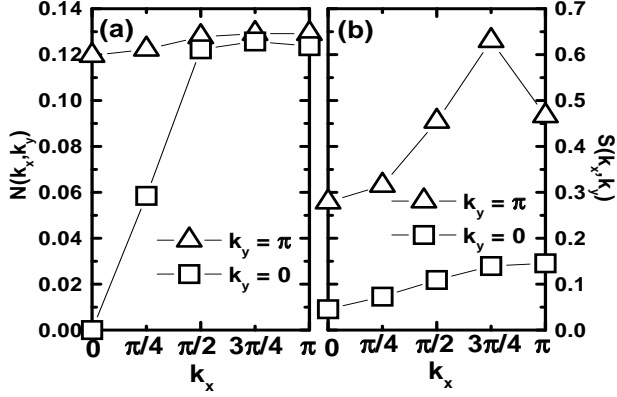,height=3.8in,width=3.25in}}
\end{center}
\vspace{-1.4in}
\caption{Results obtained with the
 DMRG technique on a 4$\times$8 cluster with four holes (x=0.125)
 at J=0.2, and t$'$=t$''$=0.0, using cylindrical boundary conditions 
(PBC along rungs).
 (a) Charge structure factor N($\rm k_x$,$\rm k_y$) vs $\rm k_x$, 
 parametric with $\rm k_y$ as indicated. 
 (b) Spin structure factor S($\rm k_x$,$\rm k_y$) vs $\rm k_x$, 
 parametric with $\rm k_y$ as indicated. Incommensurability is clear
 in the $\rm k_y$=$\pi$ branch.}
\end{figure}

\subsection{Six-leg Ladders:}

Accurate results for six-leg ladders are difficult to obtain
in the regime of parameters studied in this paper, namely small
J and nonzero t$'$-t$''$. However, some nontrivial results can still be
gathered. For instance, consider in Fig.17 the case of three-holes
on a 6$\times$4 cluster with PBC along the direction with six sites,
and OBC along the short one. In this case a good approximation to
the ground-state can be obtained using the ORBA method with
about one million 4-site rung-basis states. The procedure
to obtain the results shown in Fig.17 are the following: first it
was noticed that the central long legs are the ones the most populated by
holes. Then, one hole was projected at an arbitrary site belonging
to those central legs. In the next step, the hole density is obtained
with that hole projected, and at the position with the largest density
a second hole is now projected. The distance between the first and
second projected holes is two lattice spacings. With those holes
projected as shown in Fig.17, the hole density is recalculated. It is
clear that at two lattice spacings from the projected holes the
density has a maximum. This result is once again compatible with
a n$_h$=0.5 stripe formed this time by three-holes in a closed loop
around the direction with PBC. The stripe is not rigid but fluctuating
perpendicular to its main direction. 

\begin{figure}
\begin {center}
\mbox{\psfig{figure=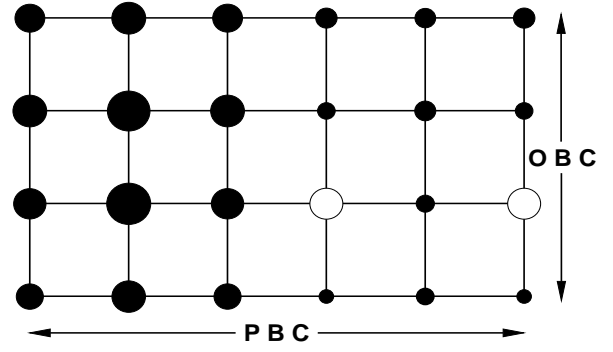,height=1.8in,width=3.0in}}
\end{center}
\caption{Results corresponding to the three holes ground-state treated with
 the ORBA technique on a 6$\times$4
 cluster with PBC in the long direction and OBC in the other, as indicated.
 The couplings are J=0.2, t$'$=-0.35, and t$''$=0.25. Two of the holes
 are projected at the open circle sites. The full circles denote the 
 probability of finding the third hole at a given site. A loop
 of mobile holes appears to form wrapping around the PBC direction. }
\end{figure}

The density of holes with a given momentum ${\bf k}$  for the
6$\times$4 cluster with three holes is shown in Fig.18. Clearly the 
momenta the most important are those around $(\pi,\pi)$, result similar
to those reported for the two-legs cluster in Fig.5b and
for a 4$\times$6 two-holes cluster with PBC in both directions in
Ref.\cite{martins2}.

\vspace{-0.1in}
\begin{figure}
\begin {center}
\mbox{\psfig{figure=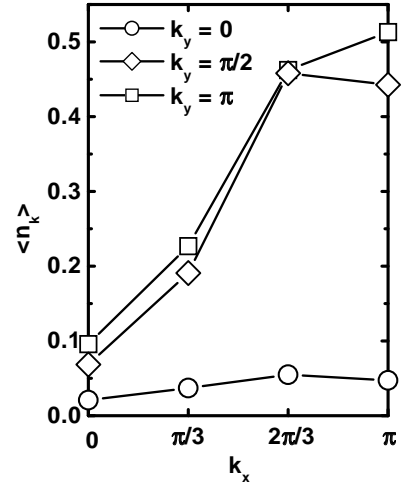,height=3.0in,width=2.3in}}
\end{center}
\caption{
n($\rm k_x$,$\rm k_y$)=$\sum_{\sigma}$$\langle$ 
 $c^{\dagger}_{{\bf k}\sigma} c_{{\bf k}\sigma}$$\rangle$ vs $\rm k_x$
 for the $\rm k_y$ momenta indicated, obtained using the ORBA
 approximation to the ground-state of three holes on a 6$\times$4
 cluster with PBC in the x-direction and OBC in the y-direction. 
$c_{{\bf k}\sigma}=\frac{1}{\sqrt{N}}\sum_{{\bf j}}e^{-i{\bf k}{\bf j}}c_{{\bf j}\sigma }$, 
where N is the number of sites in the cluster, ${\bf j}$ is a site index 
and $\sigma=+,-$.}
\end{figure}

\section{Analytical calculations: spin-polaron approach}

In previous sections, it has been shown among other items 
that the across-the-hole correlations identified before 
numerically on small square clusters for the one-hole
problem\cite{martins1} exist also on ladders. 
For completeness, in this section an analytical approach to 
the one-hole problem in two dimensions is described. 
In agreement with the above mentioned results, once again 
the existence of across-the-hole correlations is confirmed this
time using a non-numerical method, highlighting the robustness of
this feature.

The analytic approach used here
is based on the picture of the spin-bag QP or magnetic
polaron\cite{Strings,MaHo,Riera97,Vojta98}.
The spin deviations from a perfect antiferromagnet  in the vicinity 
of the hole are described by a set of fluctuation operators $A_n$
\cite{Strings,Riera97,Vojta98}, 
which create strings of spin defects attached to the hole.
The one-hole Green's function is evaluated using
a cumulant version \cite{BeckBre90} of Mori-Zwanzig projection technique.

The undoped ground state is modeled by
$|\psi\rangle = \exp(\sum_\nu \alpha_\nu S_\nu) |\phi_{\rm N{\mathaccent 19
e}el}\rangle$
where the operators $S_\nu$ create clusters of spin fluctuations
in the classical N\'{e}el state.
Following Ref. \onlinecite{SchorkFul92} it is possible to obtain a
non-linear set of equations
for the coefficients $\alpha_\nu$, which is solved self-consistently.
The fluctuation operators $A_n$ for describing the spin deviations caused
by hole motion
include the usual string operators up to a certain length $l_{\rm max}$ and
additional operators
for local fluctuation configurations.
The Green's function is calculated by projection technique in the subspace
spanned by the operators $A_n$, which amounts to the diagonalization of the
dynamical matrix $\langle\psi|A_m^\dagger H A_n|\psi\rangle$.
Processes outside the subspace formed by $A_n|\psi\rangle$ (i.e.,
self-energies) are neglected,
therefore a discrete set of poles for each spectrum is obtained.
The eigenvectors of the dynamical matrix represent the wavefunctions
corresponding
to the poles and they allow for the calculation of expectation values
such as spin correlation functions.
In the following the pole next to the Fermi surface
representing the QP is the only one considered,
the remaining part of the spectrum forms an incoherent background well
separated from the QP peak.
The approach sketched here has been successfully applied to the one-hole
problem
in several contexts \cite{Vojta98,Vojta99};
details are described in Ref. \onlinecite{Vojta98}.
In the present calculations strings with $l_{\rm max}=5$ have been used
(total number
of fluctuation operators $A_n$ around 1200).
Note that the accuracy of the approximation decreases with increasing $t$
since the employed expansion parameter is basically the amplitude of the
deviations
from the undoped state.
Therefore it is better to restrict ourselves to the parameter range with
$\rm J/t$$>$0.05
where the comparison with numerical results reveals the good accuracy of
the analytical approach.


First, let us discuss the weight of the QP pole which is known to decrease with
decreasing J/t since the size of the spin polaron increases.
The full parameter dependence of the QP weight Z at momentum $(\pi,0)$ is
shown in Fig.19a.
The introduction of t$'$$<$0 rapidly suppresses the value of 
$\rm Z_{\pi,0}$ even for intermediate or large
J, such that the entire region of moderate t$'$ is characterized by small
$\rm Z_{\pi,0}$, in good agreement 
with the computational results shown in this paper and in Ref.\cite{martins1}.
On the other hand,
the t$'$ dependence of Z at $(\pi/2,\pi/2)$ is weaker (not shown), 
also as found in Ref.\cite{martins1}. Nevertheless, its
associated QP weight Z is also strongly suppressed
with the reduction of J.

\begin{figure}
\begin {center}
\mbox{\psfig{figure=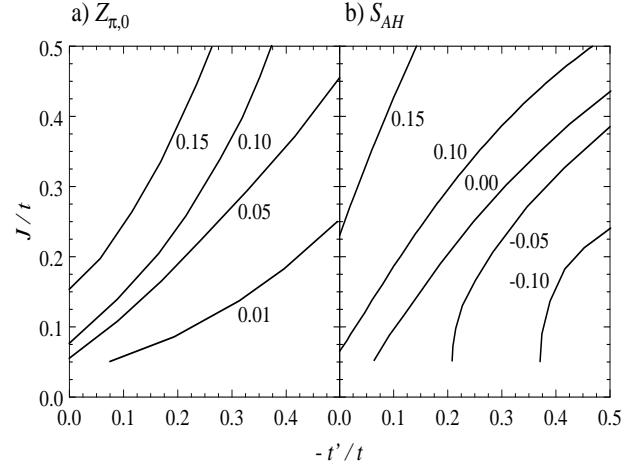,height=2.6in,width=3.2in}}
\end{center}
\caption{(a) QP weight Z and
(b) magnitude of across-the-hole correlations $\rm S_{AH}$
at momentum $(\pi,0)$ in the plane J/t vs. t$'$/t, with
t$'$/t$''$ fixed to $-1.4$.}
\end{figure}

In order to study the spin configuration near the mobile hole, 
static spin correlations in real space have been calculated relative to the
hole position.
It is known that the limit of small t (static vacancy) leads to an increase
of the antiferromagnetic correlations on bonds near the hole, whereas
hole hopping tends to ``scramble'' the spins in the hole environment which in
general weakens the AF tendencies.
This general behavior is well reproduced by our calculations.
However, the introduction of t$'$ leads additionally to antiferromagnetic
correlations $S_{AH}$ across-the-hole, see Figs.19b and 20b, quite consistent
with those 
found numerically here and in previous studies.
In Fig.20b, the strength of the antiferromagnetic bonds at
J=0.05, t$'$=-0.35, t$''$=0.25, and momentum $(\pi,0)$ are shown
as illustration.
Across the mobile hole strong antiferromagnetic correlations develop which are
supported by further strong AF bonds forming a chain segment.
The other nearest-neighbor bonds are weaker 
by a factor of two than the chain bonds.
The values of $\rm S_{AH}$, obtained with the analytic approximation used 
in this section are shown in Fig.19b. 
Although they are not as strong as found 
numerically, the qualitative trends agree quite well with computational
studies.

The stronger tendency for local spin-charge separation, at small J/t or 
in the presence of t$'$ and t$''$, can be understood within the string 
picture. The string of defect spins attached to the hole basically 
connects the spin and charge parts of the excitation, confining them 
at long distances. Therefore longer average strings correspond to less 
tightly bound spin and charge, i. e., larger spin polarons, immediately 
implying a smaller quasiparticle weight. The energy cost per unit length 
of string is proportional to J, hence 
smaller J/t allows for longer strings\cite{Strings}. Interestingly, 
the presence of t$'$ and t$''$ leads to a related effect, namely the 
hole can easily `get rid' of its attached string by intra-sublattice hopping, 
the string being then absorbed in the background spin fluctuations. This mechanism 
makes long strings effectively less costly, leading to the described large 
polarons with small QP weight Z.

\begin{figure}
\begin{center}
\mbox{\psfig{figure=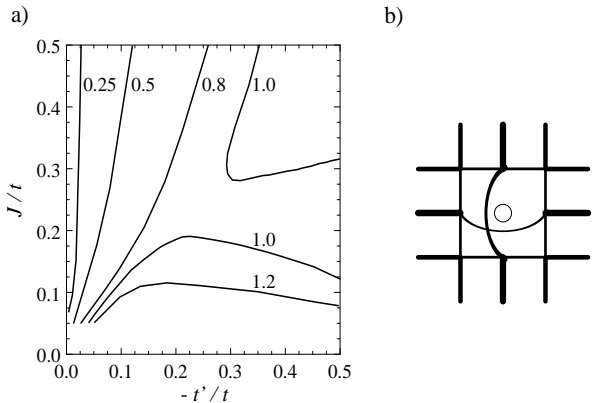,height=2.2in,width=3.2in}}
\end{center}
\caption{
(a) Ratio of the effective QP masses at momentum 
$(\pm \pi/2,\pm \pi/2)$
which is a direct measure of the (an)isotropy of the dispersion.
In the region near t$'$=0 and J/t=0.4 the value
is around 0.14 (in agreement with previous numerical 
studies [16]).
(b) Illustration of the static spin correlations near the
hole at J=0.05, t$'$=-0.35, t$''$=0.25 and momentum $(\pi,0)$.
The width of the lines is proportional to the spin-spin correlation
between sites $\rm {\bf i}$ and $\rm {\bf j}$, namely
 $-\langle {\bf S}_{\bf i}\cdot{\bf S}_{\bf j}\rangle$, and
the value of the across-the-hole correlation is -0.1.
}
\end{figure}

For completeness, in Fig.21 the calculated QP 
dispersions along high-symmetry lines
in the Brillouin zone are shown.
The points correspond to
(a) J=0.40, t$'$=0.0,
(b) J=0.40, t$'$=-0.35 and 
(c) J=0.05, t$'$=-0.35
with t$'$/t$''$=-1.4.
The change in the curvature near $(\pm\pi/2,\pm\pi/2)$
is clearly visible when adding t$'$ and decreasing J.
Whereas the dispersion is strongly anisotropic for large J and small t$'$,
there is an entire region of near isotropy for moderate values 
of t$'$ and a large range of J/t, in agreement with PES experiments\cite{arpes}.
This is also illustrated in Fig.20a where the ratio
of the effective QP masses at ${\bf K}_0$=$(\pi/2,\pi/2)$
for the entire parameter space are presented.
The masses are defined as usual as eigenvalues of the tensor ${\bf m}$
given by $\epsilon_{\bf k} = (2{\bf m})_{\alpha\beta}^{-1} (\bf k-{\bf
K}_0)_{\alpha} (\bf k-{\bf K}_0)_\beta$
where $\epsilon_{\bf k}$ is the QP energy near ${\bf K}_0$.

\begin{figure}
\begin {center}
\mbox{\psfig{figure=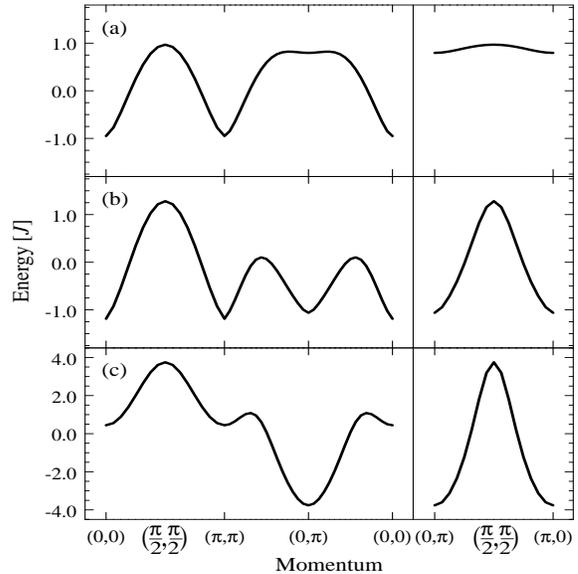,height=3.2in,width=3.2in}}
\end{center}
\caption{QP dispersions for points (a), (b) and (c) as mentioned
in the text.
The energy zero level has been set at the center of mass of the band.
The evolution towards a nearly isotropic maximum around $(\pm\pi/2,\pm\pi/2)$
is clearly visible.
}
\end{figure}

Summarizing, the introduction of t$'$ leads to a suppression
of the QP weight, especially at momentum $(\pi,0)$, a delocalization of
the spin carried by the spin polaron, and the formation of antiferromagnetic
correlations across the mobile hole. All the results are in good qualitative 
agreement with the numerical calculations reported here.

\section{Conclusions}

In this paper, tendencies toward 
spin-charge separation at short distances have been discussed in the context
of the extended t-J model using ladder geometries. 
This effort generalized previous
calculations carried out on small square clusters. Analytic
approximations have also been used, with results in good
agreement with the computational ones. Overall it is concluded
that in regimes where the hole kinetic part of the Hamiltonian
dominates, holes tend to arrange the spin environment in such
a way that across-the-hole robust antiferromagnetic correlations
are generated, both on ladders and planes. 
This arrangement helps the hole move easily among the spins,
and it resembles the structure found in one-dimensional spin-charge
separated systems. For a variety of reasons described here, it is
believed that at least at short distances
similar tendencies toward spin-charge separation are at work
in ladders and two-dimensional systems,
at very small J in the standard t-J model, or
in the extended t-J model. 
At finite hole density, holes share their nontrivial spin environment,
forming half-doped stripes as 
recently discussed by some of the authors in Ref.\cite{martins2}. 
Here more
evidence substantiating this previous result has been provided. 
The stabilization of stripe tendencies discussed here is based upon
a $small$ J/t picture, and it
has no obvious relation with those emerging in the opposite limit
of large J/t based upon the frustration of phase separated tendencies.
Although more work is certainly 
still needed to confirm the picture described here, it appears that
a gas of spinons and holons, as envisioned in two-dimensional theories
for cuprates
based upon spin-charge separation, may not be operative at low hole
densities, but instead stripes of holons seem to be forming.

\section{Acknowledgments}

E. D. thanks NSF (DMR-9814350) and the
Center for Materials Research and Technology 
(MARTECH) for support, J. C. X. thanks
FAPESP-Brazil for support and C. G. thanks 
Fundaci\'on Antorchas for partial support. 
M. V. acknowledges support by the DFG (VO 794/1-1) and 
by US NSF Grant DMR 96-23181.


\begin{references}
\bibitem{error} 
The truncation error was $\sim$$10^{-5}$
(keeping up to 800 states).
\bibitem{comm4} Results at small J and nonzero 
t$'$-t$''$ were shown in Ref.\cite{martins2} and there the inclusion of extra 
hoppings increased even further the across-the-hole correlations
along the leg direction.
\end{references}
\end{document}